\documentstyle[a4fedele,10pt]{article}

\begin{document}

\newcommand{\vm}{\vspace{0.2cm}}
\newcommand{\vl}{\vspace{0.4cm}}
\newcommand{\per}{\hspace{.2cm}}
\
\vspace{0.5cm}
\begin{flushright}
{\bf HIP-1999-60/TH}
\end{flushright}

\Large
\begin{center}
An Essay on Color Confinement

\normalsize
\vl

Masud Chaichian\\

\vm

High Energy Physics Division, Department of Physics, \\ 
University of
Helsinki\\

 and \\

Helsinki Institute of Physics\\ P.O. Box 9, FIN-00014
Helsinki, Finland\\

\vm

and

\vm

Kazuhiko Nishijima \\ 

\vm 

Nishina Memorial Foundation\\
2-28-45 Honkomagome, Bunkyo-ku, Tokyo 113-8941, Japan
\end{center}

\vspace{2.0cm}

\normalsize

\begin{center} Abstract \end{center}

\vm

Color confinement is a consequence of an unbroken non-Abelian gauge
symmetry and the resulting asymptotic freedom inherent in quantum
chromodynamics.  A qualitative sketch of its proof is presented.

\section{Introduction}

There has been an accumulation of evidence in favor of the quark model
of  hadrons [1] and we can no longer think of any other
substitute for it.  Yet, no isolated quarks have been observed to
date, and we are inclined to  think that observation of isolated
quarks is, in principle, impossible.  This is the hypothesis of quark
confinement, and it has been further extended
 to that of color confinement that implies not only the
unobservability of quarks but also of all the isolated colored particles
such as quarks and gluons.  Then a natural question is raised of
whether or not we can account for this  hypothesis within the
framework of the conventional quantum chromodynamics  (QCD) dealing
with the gauge interactions of quarks and gluons. The answer to this
question is affirmative and the detailed mathematical proof of color
confinement has been published elsewhere [2-6]. In this article,
therefore, we shall follow the flow of ideas underlying the proof in a
qualitative manner. 

The problem of color confinement may be decomposed into two steps. The
first step consists of finding a consensus of interpretations of color
confinement. Unless it is properly settled we do not know what we have
to prove in the second step. Because of the importance of this
subject many authors have proposed various interpretations. A
typical example is Wilson's area law for the loop correlation function
in the lattice gauge theory [7]. When it is obeyed the interaction
between a quark and an antiquark is given by a confining linear potential. 
Another example is given by
coherent superposition of magnetic  monopoles in the vacuum state
[8-13]. This is dual to the superconducting vacuum based on coherent
superposition of charged objects such as the Cooper pairs.
Corresponding to the superconductor of the second kind a pair of
magnetic monopoles can be connected by a quantized magnetic flux
forming a hadronic string whose energy is proportional to the
distance between them. Then the situation is similar to the preceding
example. 

In these examples one introduces a topological structure through
monopoles, strings and instantons into the configuration space. In the
present paper, however, we shall consider a different topological
structure in the state vector space. For this purpose we look for a
known example of confinement within the framework of known field
theories, and we find a prototype example in quantum electrodynamics
(QED) [2]. When the electromagnetic field is quantized in a covariant
gauge, say, in the Fermi gauge, three kinds of photons emerge,
namely, transverse, longitudinal and scalar photons, but only the
transverse photons are subject to observation leaving the other two
unobservable. We recognize that this is indeed a typical example of
confinement, and we may be able to find some clues to color
confinement by studying closely the mechanism of confinement of
longitudinal and scalar photons in QED. For this reason we analyze its
mechanism in Sec. 2 so that we can generalize it and  apply it to QCD.

One of the profound features of gauge theories is the
Becchi-Rouet-Stora (BRS) invariance [14] and its introduction is
vital to the interpretation of confinement. Therefore, we shall
describe some of the basic properties of this invariance in Sec. 3. 

The strong interactions described by QCD possess a novel feature
called asymptotic freedom [15,16], and in Sec. 4 we shall discuss how
this aspect of strong interactions drew our attention and how the
non-Abelian gauge theory entered the game. Finally in Sec. 5 we shall
combine BRS invariance with asymptotic freedom to prove color
confinement.

\section{Quantum Electrodynamics and Indefinite Metric}

When the electromagnetic field is quantized in a covariant gauge, say,
in the Fermi gauge, we find transverse, longitudinal and scalar
photons, but the latter two are never observed. We may interpret it as
an example of confinement, and we have at least three alternative ways
of explaining it. First, we can refer to the representations of the
Poincar\'{e} group for massless particles [17,18]. Then, massless particles
are known to have only two directions of polarization no matter what
their spin is. Thus photons are always transversely polarized and the
same would be true with gluons if they could be observed. The second
method is to employ the Coulomb gauge by keeping only the transverse
photons from the start. The remnants of unobservable  photons manifest
themselves in the form of the Coulomb potential. This method is
applicable, however, only to the linear Abelian gauge theories such as
QED. The third and the most useful method is the introduction of a
subsidiary condition such as the Lorentz condition.

Quantization of the electromagnetic field in a covariant gauge forces
us to introduce indefinite metric [19] which is inherited from the
Minkowski metric. Thus the whole state vector space in QED can no
longer possess the positive-definite metric, and for the physical
interpretation of the theory we have to eliminate indefinite metric by
imposing the Lorentz condition on the state vectors to select
observable or physical states. In order to execute this program let
us quantize the free electromagnetic field in the Fermi gauge and for
a given momentum we have four directions of polarization, namely, two
transverse, one longitudinal and one scalar. Thus we have four kinds of
photons specified by the directions of polarization. The canonical
quantization then implies that the scalar photons are represented by
negative norm states. This is a consequence of the manifest covariance
of the quantization of the vector field in the Minkowski space.

The emergence of indefinite metric indicates that observable states
occupy only a portion of the whole state vector space called the
physical subspace. In order to define such a subspace we introduce a
subsidiary condition known as the Lorentz condition. Let us consider
the four-divergence of the vector field, then it represents a free
massless field even in the presence of the interactions. We decompose
it into a sum of positive- and negative- frequency parts
corresponding to destruction and creation operators, respectively. We
find that the photons involved in this operator are special
combinations of the longitudinal and scalar photons in the
amplitude. We shall call them a-photons, then an a-photon state has
zero norm. We can introduce an alternative combination of longitudinal
and scalar photons called b-photons in such a way that a b-photon
state also has zero norm. Thus for  a given momentum we have two
transverse (t-) photons, an a-photon and a b-photon. Although both
an a-photon state and a b-photon state have zero norm, their inner product
is non-vanishing so that they are metric partners. 

A physical state is defined as such a state that is annihilated by
applying the positive frequency part of the four-divergence of the
vector field. This is the Lorentz condition. We can easily verify that
the S matrix in QED transforms a physical state into another physical
state since it commutes with the four-divergence. This is one of the
general features of the subsidiary condition. Also we can easily
verify that the b-photons are excluded from the physical
subspace. Therefore, we have only t-photons and a-photons in the
physical states.  Then we can show that the  inner product of a physical
state involving at least one a-photon  with another physical state
vanishes identically. In other words, a-photons give no contributions
to observable quantities, and both a- and b-photons escape
detection. This is the confinement mechanism of the longitudinal and
scalar photons. In QED only the transverse photons remain observable.
In QCD, however, not only longitudinal and scalar gluons but also
transverse gluons are  unobservable. Thus, there are some essential
differences in the nature of confinement between QED and QCD. In the
former case confinement is kinematical in the sense that it could be
understood without recourse to dynamics of the system, whereas in the
latter case it is dynamical in nature as the proof depends sensitively on
the dynamical properties of the system.

\section{Quantum Chromodynamics and BRS Invariance}

As we shall see in the next section strong interactions of quarks are
mediated by a non-Abelian gauge field corresponding to the  $SU(3)$
 color
symmetry. Thus we shall discuss one of the most characteristic features of
gauge theories known as the BRS invariance in this section [14].

Classical electrodynamics is gauge-invariant. Field strengths
expressed in terms of the vector  field are invariant under the local
or space-time-dependent gauge transformations of the latter. Given a
source term, therefore, the solution of the equation for the vector
field is not uniquely given, and this non-uniqueness is an obstacle
to quantization. In order to overcome this difficulty we add to the
gauge-invariant  Lagrangian a term  violating the local gauge
invariance. This extra term is called the gauge-fixing term and was
first introduced by Fermi. Later it has been generalized so as to
include an arbitrary parameter called the gauge parameter. In the
original form introduced by Fermi this parameter is equal to unity. 

After quantization we find that we have to introduce indefinite metric
into the state vector  space and that the divergence of the vector
field commutes with the S matrix.  Because of the inclusion of the
gauge-fixing term the field equation deviates from the classical
Maxwell equation by a term proportional to the four-divergence of the
vector field. It so happens that a matrix element of this
four-divergence between two physical states vanishes identically
because of the Lorentz condition, and the classical Maxwell equation is
recovered in the physical subspace. In this way we find, despite the
introduction of the gauge-fixing term, that expectation values of
gauge-invariant quantities and the S matrix elements in the physical
subspace are independent of the choice of the gauge parameter
because of the congeniality between the gauge-fixing term and the
subsidiary condition. In what follows we shall extend this approach to
QCD. 

There are many essential differences between QED and QCD, however.
The former is an Abelian gauge theory described by a linear field
equation, whereas the latter is a non-Abelian gauge theory described
by a non-linear field equation. In both cases the gauge-invariant
part of the Lagrangian is given by  the square of the field
strength. So, let us introduce the gauge-fixing term in QCD assuming
the same structure as in QED. Then we recognize that it does not work
because observable quantities depend explicitly on the gauge
parameter. Another difficulty arises from the fact that the
four-divergence of the gauge field is no longer a free field, and this
prevents us from defining its positive frequency part. In other
words, the Lorentz condition cannot be employed to define physical
states in QCD. Thus we are obliged to find a device to overcome these
difficulties and to this end we shall introduce the Faddeev-Popov
ghost fields. 

In order to eliminate the gauge-dependence of physically relevant
quantities Faddeev and Popov have proposed a procedure of averaging the
path integral over the manifold of gauge transformations. We skip the
mathematical detail here and refer to the original paper [20], but we
should mention that this procedure resulted in a new additional term in
the Lagrangian called the Faddeev-Popov (FP) ghost term. This term
involves a pair of Hermitian scalar fields, but they are anticommuting
and consequently violate Pauli's theorem on the connection between
spin and statistics.  For this reason they are called ghost
fields. Pauli's theorem is based on three postulates, (1) Lorentz
invariance, (2) local commutativity or microscopic causality and (3)
positive-definite metric for state vectors, and the FP ghost fields
violate the last one obliging us to introduce indefinite metric into
the theory.

Thus we face again the problem of eliminating indefinite metric from
the theory with the help of an appropriate subsidiary condition to
select physical states out of the  whole state vector space. When
physical states are so defined  as those that are annihilated by applying
a certain operator, that operator should commute with the S matrix
as does the four-divergence of the vector field in QED. In order to
find such an operator a novel symmetry discovered by Becchi, Rouet and
Stora is extremely useful. Although this symmetry was originally
utilized in renormalizing QCD, it plays an essential role in the proof
of color confinement in QCD. 

In a classical gauge theory a local gauge transformation is specified
by a function of the space-time coordinates called the gauge function
and the classical theory is invariant under such a
transformation. This local gauge invariance is lost when the
gauge-fixing and FP ghost terms are introduced. Besides, local gauge
transformations are defined only for the color gauge field and the
quark fields, but they are not even defined for FP ghost fields.  The
BRS  transformations for the color gauge field and the quark fields are
given by replacing the gauge function by one of the FP ghost fields in
infinitesimal gauge transformations. Since we have a pair of ghost
fields we introduce, correspondingly, a pair of BRS
transformations. Then a question is raised of how to define BRS
transformations of the ghost fields since their gauge transformations
are not defined. Fortunately, this problem has a simple but beautiful
solution. Their BRS transformations are introduced by demanding the
invariance of the total Lagrangian  under them. 

The total Lagrangian including the gauge-fixing and FP ghost terms is
no longer invariant under local gauge transformations, but it 
is invariant under the global BRS transformations. Noether's
theorem then tells us that there must be a pair of conserved
quantities corresponding to a pair of BRS symmetries. They are
Hermitian and called the BRS charges. As mentioned before there  are
two kinds of Hermitian FP ghost fields and correspondingly a BRS
charge must involve one of the ghost fields. In what follows we keep only
one of these two charges for simplicity. The BRS charge that we keep
is anticommuting just as the FP ghost field, and consequently the
square of the BRS charge vanishes and it is called nilpotent. The
Hermiticity and nilpotency of the BRS charge would imply indefinite
metric since otherwise it would be a null operator [21,22]. The
nilpotency is important and allows us to introduce the concept of
cohomology in the theory. After a long detour we are going to introduce
an appropriate subsidiary condition. Physical states are defined as
those states that are annihilated by applying the BRS charge [23].

The FP ghost fields do not appear in the conventional QED but we can
also introduce them although they are non-interacting fields. Then we
can combine the Lorentz condition with the additional condition
implying the absence of FP ghosts to define the physical states. When
these conditions are satisfied, we can prove that physical states so
defined are annihilated by the BRS charge in QED. 

The BRS charge is the generator of the BRS transformation and the BRS
transform of an operator is given by the commutator or anticommutator of
that operator with the BRS charge, and this transformation is also
nilpotent. An operator which is the BRS transform of another operator
is called an exact operator, then it is clear that the matrix element of
an exact operator between a pair of physical states vanishes. 

The equation for the non-Abelian gauge field deviates from the
classical Maxwell equation  and in fact the divergence of the field
strength plus the color current does not vanish but is equal to a
certain exact operator, which will be referred to as an exact current
hereafter. Therefore, the classical Maxwell equation is recovered when
we take  the matrix element of the field equation between a pair of
physical states. Furthermore, the BRS charge commutes with the S
matrix. Thus the scenario in QED is reproduced almost exactly.

When single quark states and single gluon states are unphysical these
particles are unobservable and consequently confined. Thus the problem
of color confinement reduces to that of proving that they are
unphysical states. We shall evaluate the expectation value of the
exact current in a single quark state or a single gluon state. If they
should belong to physical states the expectation values in these
states would vanish identically, so that non-vanishing of the
expectation values would be a direct indication that these particles
are unphysical and confined.

The four-divergence of the exact current vanishes, and we can give a
set of Ward-Takahashi identities for Green's functions involving the
exact current [2-4]. By making use of the above set of Ward-Takahashi
identities we can prove that the expectation value of the exact
current in a single colored particle state survives when the exact current
as applied to the vacuum state does not generate a massless spin zero
particle. Therefore, the absence of such a massless particle is a
sufficient condition for color confinement [2-4]. In order to check
its absence we introduce the vacuum expectation value of the
time-ordered product of the gauge field and the exact current and
evaluate the residue {\it C} of the massless spin zero pole of the
Fourier transform of this two-point function. The four-divergence of
this two-point function is proportional to this constant {\it C}
except for a trivial kinematical factor, and the divergence can be
cast in the form of an equal-time commutator. 

By checking this equal-time commutator closely we find that {\it C} is
the sum of a constant {\it a} and the Goto-Imamura-Schwinger (GIS)
term. The constant {\it a} is equal to the inverse of the
renormalization constant of the color gauge field. These constants
{\it C} and {\it a} satisfy  distinct renormalization group (RG)
equations and boundary conditions.  We shall not enter this subject
here since the mathematical detail has been given elsewhere [2-5], but
we infer the fact that vanishing of {\it a} automatically leads to
vanishing of {\it C} and color confinement is realized. Indeed, it has
been known for some time that gluons are confined when {\it a}
vanishes [24,25], but now with the help of the BRS invariance we
could conclude that not only gluons but also all the colored particles
are simultaneously confined. We shall come back to this subject again
in Sec. 5.

\section{Asymptotic Freedom}

In this section we shall review briefly how and why our attention was
drawn to the non-Abelian gauge theory in describing strong
interactions. In particle physics strongly interacting particles such
as nucleons and pions are called hadrons. Hadrons are composite
particles of quarks and antiquarks, however, and we have to study
the origin of the strong interactions of quarks. 

We already know that strong interactions are mediated by the color
gauge field and the quanta of this field are called the gluons since
they glue up quarks together to form hadrons. Dynamics of quarks and
gluons is called QCD as mentioned before. In the sixties experiments on
the deep inelastic scattering of electrons on protons had been carried
out. The differential cross-section had been measured by specifying
the energy and direction of electrons without observing the hadrons
in the final states. Then, apart from kinematical factors this
differential cross-section can be expressed as a linear combination of
two structure functions. They are functions of the square of the
momentum transfer and the energy loss of the electron in the
laboratory system. When these two variables increase indefinitely the
two structure functions tend to be functions of the ratio of these two
variables except for trivial kinematical factors. This characteristic
behavior of structure functions is called the Bjorken scaling [26],
and it is considered to be an empirical manifestation of the
properties of strong interactions. What do we learn from this?  In
1969 Feynman proposed the parton model and assumed that a nucleon
consists of point-like partons moving almost freely inside the nucleon
[27]. In order to keep the partons inside the nucleon, however, the
four-momentum of a parton must be equal to a
fraction {\it x} of the total four-momentum of the nucleon. The
partons may be identified with the quarks and since {\it x} is
identified with the ratio of the two kinematical variables referred to
in the above the distribution of the fraction {\it x} has been shown
to be related to the structure functions.

>From the success of the parton model in reproducing the Bjorken
scaling we may infer that quarks inside the hadrons are almost free
and that the interactions of quarks turn out to be weaker at shorter
distances. This is a distinctive feature of strong interactions and we
may express it in the momentum space as follows: The probability of a
process involving large momentum transfer in strong interactions is
small. 

We look for a model satisfying this condition and find that only
non-Abelian gauge interactions meet this requirement with the help of
RG [15,16].

The concept of RG was first introduced by Stueckelberg and Petermann in
1953 [28], and it was further advanced by Gell-Mann and Low in QED in
1954 [29].  Let us consider a dielectric medium and put a positive
test charge inside, then the medium is polarized, namely, negative
charges are attracted and positive ones are repelled by this test
charge. As a consequence it  induces a new charge distribution in the
medium. The total charge inside a sphere of radius {\it r} around the
test charge is a function of {\it r} and we call it the running
charge. The vacuum is an example of the dielectric media because of
its ability of being polarized -- the vacuum polarization.  In  this
case the test charge is called the bare charge and the total charge
inside a sphere of a sufficiently large radius is called the
renormalized charge. The running charge is a function of the radius
{\it r}, but it can also be regarded as a function of momentum
transfer through the Fourier transformation. The bare charge then
corresponds to the limiting value of the running charge for infinite
momentum transfer. 

Gell-Mann and Low have proved on the basis of the RG method that given
a finite renormalized charge the bare charge is equal to a certain
finite constant independent of the value of the renormalized one or
it is divergent [29].  The Bjorken scaling phrased in terms of RG
implies that the bare coupling constant must be equal to zero. We
shall refer to this property as asymptotic freedom (AF), and the
non-Abelian gauge theory is the only known example in which AF is 
realized as clarified by Gross and Wilczek and by Politzer [15,16].
The origin of AF may be traced back to the fact that the vector field
introduces indefinite metric needed to realize AF and that the
non-Abelian gauge theory is the only example involving non-linear
interactions of the vector field.

Thus starting from the empirical Bjorken scaling we have finally
reached the non-Abelian gauge theory of strong interactions, namely,
QCD.

\section{Color Confinement}

Now we are ready to present the proof of color confinement, at least
verbally, by combining arguments given in preceding sections.

In QED the square of the ratio of the renormalized charge to the bare
one is equal to the renormalization constant of the electromagnetic
field. It is equal to the inverse of the dielectric constant of the
vacuum relative to the empty geometrical space.  Usually the
dielectric constant of a dielectric medium is defined relative to the
vacuum, but here we define it relative to the empty geometrical space
or the void. 

This dielectric constant of the vacuum is larger than unity as a
consequence of the positive-definite metric of the physical subspace,
or more intuitively, it is a consequence of the screening effect due
to the vacuum polarization. Then, let us consider a fictitious case in
which the dielectric constant of the vacuum is smaller than unity. In
this case we have antiscreening instead of screening when a test
charge is placed in this fictitious vacuum, and such a vacuum is
realized when a pair of virtual charged particles of indefinite metric
should contribute to the vacuum polarization. In this case the
running charge would be an increasing function of the radius {\it r}
at least for small values of {\it r}. Next we shall consider an
extreme case of the vanishing dielectric constant, then a small test
charge would attract an unlimited amount of like charges around it
thereby bringing the system into a catastrophic state of infinite
charge. Nature would take safety measures to prevent such a state from
emerging, and a possible resolution is to bring another test particle
of the opposite charge. The total charge of the whole system is
equal to zero and charge confinement would be realized. In QED,
however the dielectric constant of the vacuum or the inverse of the
renormalization constant is larger than unity, and the above scenario
reduces to a mere fiction. 

The situation in QCD is completely different since it allows
introduction of indefinite metric in the vacuum polarization and AF
is one of its manifestations. In QCD  what corresponds to the
dielectric constant of the vacuum in QED is the inverse of the
renormalization constant of the color gauge field denoted by {\it a}
in Sec. 3. If {\it a} should vanish we would encounter a scenario
similar to the one mentioned above and a test color charge would
induce an intolerable catastrophic state. In Sec. 3 we have shown that
such a state is excluded by means of the subsidiary condition that
selects physical states. Therefore, what can be realized are states
of zero color charge and this is precisely color confinement. Unlike
electric charge, color charge is not a simple additive quantum number
but a member of a Lie algebra $su(3)$, so that physically realizable
states should belong to the one-dimensional representation of this
algebra. Thus the entire problem of color confinement reduces to the
proof that the constant {\it a} vanishes.

Before presenting its proof we have to introduce the concept of the
equivalence class of gauges [2,4,5]. When the difference between two
Lagrangian densities is an exact operator we say that these two
Lagrangian densities belong to the same equivalence  class of
gauges. For instance, two Lagrangian densities corresponding to two
distinct values of the gauge parameter belong to the same equivalence
class. In QCD hadrons are represented by BRS invariant composite
operators [30-32], and the S matrix elements for hadron reactions are
obtained by applying the reduction formula of Lehmann, Symanzik and
Zimmermann [33] to Green's functions defined as the vacuum
expectation values of the time-ordered products of the BRS invariant
composite operators. Then we can readily prove that the S matrix
elements for hadron reactions are the same within the same equivalence
class of gauges [2,4,5]. Color confinement signifies that the
unitarity condition for the S matrix in the hadronic sector is
saturated by hadronic intermediate states. That means that quarks and
gluons have no place to show up in the unitarity condition just as
longitudinal and scalar photons never appeared in the S matrix
elements in QED. Therefore, we may take it for granted that the
concept of color confinement is gauge-independent within the same
equivalence class.

Then we come back to the evaluation of the constant {\it a}. First, it
should be stressed that {\it a} can be evaluated exactly as a function
of the gauge coupling constant and the gauge parameter thanks to AF
[2,5].  These two parameters define a two-dimensional parameter space,
which is then decomposed into three domains according to the value of
{\it a}, namely, zero, infinity and finite. It should be stressed
here that the existence of these three domains can be proved without
recourse to perturbation theory.  Of these three domains color
confinement is manifestly realized in the first one, and also in the
other two confinement should prevail because of the gauge-independence
of the concept of color confinement. Evaluation of {\it a} by means of
RG based on AF is a very interesting mathematical problem, but we
shall refer to the original paper for the technical detail[5].

Finally, it should be stressed that confinement as has been discussed
in this paper is realized only when we have an unbroken non-Abelian
gauge symmetry [2]. When a certain gauge symmetry is spontaneously
broken the exact current generates a massless spin zero particle as
the Nambu-Goldstone boson and our proof of confinement breaks down.  For
instance, the electroweak interactions are formulated on the gauge
group  $SU(2)\times U(1)$, but spontaneous symmetry breaking reduces
the gauge symmetry to the Abelian $U(1)$  corresponding to the
electromagnetic gauge symmetry. Thus the electroweak interactions do
not possess any unbroken non-Abelian gauge symmetry and are not
capable of confining any particle. 

To conclude, we have presented the flow of ideas towards intuitive
understanding of the mechanism of color confinement without recourse
to mathematical detail, but interested readers  are encouraged to
refer to the original articles. 

Acknowledgement: The financial support of the Academy of Finland under 
the Project no. 163394 is greatly acknowledged.
The authors are grateful to Professor A. N. Mitra for kindly inviting
us to contribute this article to the INSA book.

\end{document}